\documentclass[twocolumn,showpacs,amsmath,amssymb,prl]{revtex4}%
\usepackage{graphicx}% Include figure files
\usepackage{dcolumn}% Align table columns on decimal point
\usepackage{bm}% bold math

\def\beq{\begin{equation}}
\def\eeq{\end{equation}}
\def\beqa{\begin{eqnarray}}
\def\eeqa{\end{eqnarray}}
\newcommand{\beqas}{\begin{eqnarray*}}
\newcommand{\eeqas}{\end{eqnarray*}}
\def\nat{\mbox{nat}}

\def\Rb{{\bf R}}

\def\<{\langle}
\def\>{\rangle}

\newcommand{\cut}[1]{}

\cut{
\setlength{\textheight}{23.4cm}
\setlength{\topmargin}{-0.75cm}
\setlength{\textwidth}{15.5cm}
\setlength{\oddsidemargin}{0.5cm}
}

\begin{document}

\preprint{APS/123-QED}

\title{Teaching computers to fold proteins}

\author{Ole Winther}
\email{owi@imm.dtu.dk}
\altaffiliation[Present address:]{
Informatics and Mathematical Modelling, Technical
University of Denmark, 2800 Lyngby, Denmark.}
\author{Anders Krogh}
\email{krogh@binf.ku.dk}
\altaffiliation[Present address: ]
{Bioinformatics Centre, University of Copenhagen,
Universitetsparken 15, 2100 Copenhagen, Denmark.
}
\affiliation{Center for Biological Sequence Analysis,
The Technical University of Denmark,
Building 208, DK-2800 Lyngby, Denmark}

\date{\today}% It is always \today, today,
             %  but any date may be explicitly specified

% PACS, the Physics and Astronomy
                             % Classification Scheme.
%\keywords{Suggested keywords}%Use showkeys class option if keyword
                              %display desired
\begin{abstract}
A new general algorithm for optimization of potential functions for
protein folding is introduced. It is based upon gradient optimization of
the thermodynamic stability of native folds of a training set of proteins
with known structure. The iterative update rule contains two thermodynamic
averages which are estimated by (generalized ensemble) Monte Carlo.
%The first average
%over the native conformational space is stabilizing the native fold,
%whereas the second average over the entire conformational space 'unlearns'
%incorrect folds.
We test the learning algorithm on a Lennard-Jones (LJ)
force field with a torsional
angle degrees-of-freedom and a single-atom side-chain.
%The tunable parameters
%are the interaction strengths, characteristic LJ radii and $C_\alpha$ to side
%chain distances.
In a test with 24 peptides of known
structure, none folded correctly with the initial potential functions,
but two-thirds came within 3{\AA} to their native fold after
optimizing the potential functions.
% A detailed thermodynamic analysis of the results suggests that the
% simplified potential is inadequate for protein folding, We discuss the
% choice of potential functions, restrictions of the conformational search space and
% computational issues regarding
% scaling the method up to training on ${\cal O}(10^2)$ short proteins.
\end{abstract}

\maketitle

%\section*{Introduction}

It is one of the long-standing challenges of science to simulate
protein folding in a computer and predict the three-dimensional
structure -- the native fold.  According to Anfinsen's hypothesis the
native fold of a protein is the one with the lowest free energy
\cite{An73}.  To fold a protein in silico, it is therefore necessary
to have a sufficiently good description of the energetics of the
system.
Even the most sophisticated all-atom potentials
\cite{BrBrOlStSwKa83,MoMcBuSc75} and statistical potential functions
\cite{MiJe96,SaMo98,Si90} will not usually give stability of an
experimentally determined native structure. Furthermore, these potential
functions have so many degrees of freedom that nano-second time-scale
molecular dynamics simulations require of the order of months on even
the fastest computers.  To sample the state space of a protein in
solution with present-day computers it is therefore necessary to use a
simplified description of the protein and the solvent rather than an
all-atom model. It is virtually impossible to calculate such
potential functions from first principles.
%AK deleted 2/9/03
\iffalse
Only a few years back it was predicted that this kind
of \emph{ab initio} structure prediction would be impossible in the
foreseeable future \cite{Le97}. However, a recent growth in activity and
the introduction of a blind prediction exercise
\cite{CASP} have shown that it is indeed possible to predict
protein structure with a root mean squared deviation (RMSD)
of 4-6{\AA} with some methods, see e.g.\ \cite{BaCASP4,ChRoBoBa03}.
These new methods have yet to reach their full potential and
it is likely that the coming years will see improvements
spurred by developments in potential functions, search algorithms,
sophisticated use of database information and computer technology.
\fi

In this paper we describe a method to estimate parametrized potential
functions from a training set of known protein structures.
Most previous work on
estimation of potentials use statistical approaches
\cite{MiJe96,SaMo98,Si90}, which are
based on static structures.  The main new feature in our approach is
that we optimize the potentials during simulation of the folding
process, so as to maximize the thermodynamic probability of
the native folds of the whole training set.
This maximum likelihood estimation procedure, which is essentially
Boltzmann learning \cite{HiSe83}, can be thought of as iteratively
stabilizing the native structure on the one hand and 'unlearn'
incorrect folds, which traps the protein during folding, on the other.
There exists other approaches that are similar in spirit, but none which
aim directly at optimizing the thermodynamic stability. Rather,
related measures are optimized such as
the normalized difference between the native energy and the
average energy over all alternative conformations
\cite{GoLu-ScWo92,HaSc96,MiSh96}, the thermodynamic average of the
overlap to the native state in a contact energy model
\cite{SeVeMaBa96,BaVeKn00,BaFaKnVe01}
and linear optimization methods for ensuring that the native state has
lowest free energy
\cite{RoPhOhDi00,MiSeBaMa01}. The overlap method is the one closest related to
optimizing the thermodynamic stability.

%AK deleted 2/9/03 %%%%%%%%%%%%%%%%%%%%%%%%%%%%%%%%%%%%%%%%%%
\iffalse
In a test we optimized a simplified force field potential
on a set of 24 peptides
of known structure.  None of these folded correctly with the initial
potential functions, but two thirds came within 3{\AA} of
their native fold after optimizing the potential functions.

%\section*{Algorithm and Model}

In the following we describe the algorithm for optimizing the thermodynamic stability of the
native structure of a training set of protein sequences with known native fold
and thereafter give the protein model and energy function used for testing the algorithm.
\fi               %%%%%%%%%%%%%%%%%%%%%%%%%%%%%%%%%%%%%%%%%%

In the general setup we have a parameterized energy function $E_\theta(\Rb,\mbox{seq})$ with
parameters $\theta$, which give the energy for an amino acid sequence
$\mbox{seq}$ with atomic coordinates $\Rb$.
The probability of finding the $i$th training sequence
in its native state is given by the Boltzmann weighted volume of conformation
space compatible with
 the {\it native structure\/} divided by the {\it total\/}
Boltzmann weighted volume of conformation space
\beq
P(\nat_i|\mbox{seq}_i,\theta) =
\frac{\int_{\nat_i} \exp(-\beta E_\theta(\Rb,\mbox{seq}_i))\;d\Rb}
      {\int          \exp(-\beta E_\theta(\Rb,\mbox{seq}_i))\;d\Rb}  ,
\eeq
where $\beta=1/kT$ and the integral in the numerator is only over the part of
conformation space associated with the native structure. The definition and choice of the size of the
native volume in conformation space should reflect all expected variability such as
the loss of description accuracy due to the crudeness of the protein model,
thermal variability of the native state and the uncertainty in the determination of
the crystal/NMR structure.
% The two latter factors are expected to play a smaller role
% when using pdb protein structures, but may be important for sequences with a
% less well determined native structure such as the peptides used in our study.
%In this study we use a hypersphere with a radius of 1{\AA} around the
%$C_\alpha$ coordinates of the crystal structure.
In this study we define the native volume as all structures within
a $C_\alpha$ root mean square deviation (RMSD) of 1\AA\ from the
crystal structure.
%Instead of a fixed volume it would also be possible to use
%a weighting function which decreases with the distance to the
%crystal structure.

The
%parameters $\theta$ are the same for all training sequences and the
objective is to maximize the joint probability $\prod_{i=1}^N
P(\nat_i|\mbox{seq}_i,\theta)$ with respect to the parameters
$\theta$, where $N$ is the number of sequences in the training
set. We choose to perform the maximization by gradient ascent
$\theta^{\rm new} := \theta^{\rm old} + \Delta \theta$, where
\beqa\label{Boltzmann_update} \Delta \theta &=& \eta \nabla_\theta
\sum_i \ln P(\nat_i|\mbox{seq}_i,\theta)
\\
&=& \eta \beta  \sum_i
\left[  \< \nabla_\theta E_\theta(\Rb,\mbox{seq}_i)  \> - \<
        \< \nabla_\theta E_\theta(\Rb,\mbox{seq}_i)  \>_{\nat_i}
\right] \nonumber
\eeqa
with $\eta$ being the learning rate, $ \< \ldots \>$ and $ \< \ldots  \>_{\nat_i}$ denoting
Boltzmann averages over the total conformation space and the part
associated with the native structure, respectively. In neural
computation context this is known as the Boltzmann learning rule
\cite{HiSe83}. In simulations we perform the Boltzmann averages by
a (generalized ensemble) Monte Carlo method.

% where the average over the native volume is simply defined as the
% conformations within a certain RMSD cut-off (here 1 \AA)
% from the native structure, see below.

The above learning rule applies to any differentiable potential function.
The aim is to estimate
a potential that gives a high probability to the correct fold
under the given protein model and with the chosen simulation
method, which of course does not
guarantee that the potential is close to the real physics.
To demonstrate the validity of the method we
have applied it to a simple force field.
% The limitations of the solution found by the learning algorithm
% for this potential point to possible modifications, see the Discussion.
The amino acid unit model has 6 atoms
{\small $- \mbox{\hspace*{-1.5mm}}
   \begin{array}[t]{c}
     N \\ H
   \end{array}
 \mbox{\hspace*{-2mm}} \begin{array}[b]{c}
      R_i  \\ C_\alpha
   \end{array}
 \mbox{\hspace*{-2mm}} \begin{array}[t]{c}
      C' \\ O
   \end{array} \mbox{\hspace*{-2mm}}  -
$},
where $O$ and $H$ is introduced to be
able to define backbone hydrogen bonds and to give a more realistic local torsion potential.
The whole side chain is represented by $R_i$, so the parameters relating to
this is amino acid dependent, whereas the other atoms are treated more
conventionally. The conformational degrees of freedom are the torsional angles
$\phi$ and $\psi$ (rotation around the $N C_\alpha$ bond and $C_\alpha C'$ bond).
The $C_\alpha R_i$ distance is adaptive (one parameter for each of the 20 amino acids), whereas all
other bond lengths and angles are fixed to their average value as given in Ref.\ \protect\cite{Cr92}.
The angle to $R_i$ is fixed to the average value for
$C_\alpha C_\beta$.
\iffalse % removed by Ole 3-9
The amino acid model, shown schematically in Figure \ref{fig:backbone}, has
a realistic backbone with the torsional angles
$\phi$ and $\psi$ as
the conformational
degrees of freedom and a simplified one atom side-chain.
This
gives 6 atoms per amino acid where $O$ and $H$ is introduced to be
able to define backbone hydrogen bonds and to give a more realistic local torsion potential.
The whole side chain is represented by $R_i$, so the parameters relating to
this is amino acid dependent, whereas the other atoms are treated more
conventionally.
\fi %%%%%%%%%%%%%%%%%%%%%%%%%%%%%%

The energy is a generalization of the one proposed
in Ref.\ \cite{IrSjWa00} which in turn is inspired by the classical force fields.
The energy is split into local and non-local terms
$
E= E_{\rm local} + E_{\rm non-local}
$.
The local interactions are mainly introduced to model
local steric constraints and the non-local consists of
three types of terms
introduced to model pairwise interactions, hydrophobic,
surface and related effects (hp) and hydrogen bonding (hb):
$ E_{\rm non-local} =  E_{\rm hp} +  E_{\rm hb}$.

% The
% specific terms in the potential---of which the most important are
% of the  Lennard Jones type---are described in the Appendix.

%\section*{Appendix}

% In the following the potential function is described in detail.
%\paragraph{Local energy.}

The local energy contains two types of terms
$
E_{\rm local} = \frac{\epsilon_{\phi}}{2} \sum_i (1+\cos 3 \phi_i)
               +\frac{\epsilon_{\psi}}{2} \sum_i (1+\cos 3 \psi_i) +
%\lefteqn{
\sum_i \sum_{X,Y} \epsilon_i^{XY}
\left\{
  5  \left(  \frac{\sigma^{XY}}{r_{i}^{XY}} \right)^{12}
- 6 \left( \frac{\sigma^{XY}}{r_{i}^{XY}} \right)^{6}
\right\}
%}
$.
\iffalse
\beqa
E_{\rm local} &=& \frac{\epsilon_{\phi}}{2} \sum_i (1+\cos 3 \phi_i)
               +\frac{\epsilon_{\psi}}{2} \sum_i (1+\cos 3 \psi_i) +
\nonumber \\
\lefteqn{
\sum_i \sum_{X,Y} \epsilon_i^{XY}
\left\{
  5  \left(  \frac{\sigma^{XY}}{r_{i}^{XY}} \right)^{12}
- 6 \left( \frac{\sigma^{XY}}{r_{i}^{XY}} \right)^{6}
\right\}.
}
\eeqa
\fi
The sums on $i$ are over the individual amino acids.
The first two terms -- which are probably not so important -- introduce
a three fold symmetry and has two parameters independent of the side chains.
The last term is mainly introduced to model steric
constraints, and the sum over $X,Y$ runs over a restricted set of pairs
 of local atoms in amino acid $i$, $i+1$ and $i+2$
(with terms set to zero when $i+1$ or $i+2$ are
larger than the length of the protein). The
following atom pairs are included:
$H(i)R(i)$, $R(i)O(i)$, $H(i)H(i+1)$,
$O(i)H(i+2)$ and $O(i)O(i+1)$. Only the first two pairs depend on the
amino acid side chain, so this introduces a total of 2*(20+20+2+2+2)=92
parameters.

%\paragraph{Hydrogen bonding}
Hydrogen bonding is only considered for
backbone $NH$ and $C'O$ pairs.
The hydrogen bond energy contains
both angle dependence %\cite{DaGoMa97}
and a 12--10 Lennard-Jones potential
with two adjustable parameters
$
E_{\rm hb}  = \epsilon_{hb} \sum_{ij} u_{ij}
%u(\theta_{ij}^{NHO},\theta_{ij}^{HOC'})
\left\{
\left( \frac{\sigma_{hb}}{r_{ij}^{HO}} \right)^{\!\!12}
\!\!\!\!
- 2 \left( \frac{\sigma_{hb}}{r_{ij}^{HO}} \right)^{\!\!10}
%\bigg
\right\}
$,
\iffalse
\begin{equation}
E_{\rm hb}  = \epsilon_{hb} \sum_{ij} u_{ij}
%u(\theta_{ij}^{NHO},\theta_{ij}^{HOC'})
\bigg\{
\bigg( \frac{\sigma_{hb}}{r_{ij}^{HO}} \bigg)^{\!\!12}
\!\!\!\!
- 2 \bigg( \frac{\sigma_{hb}}{r_{ij}^{HO}} \bigg)^{\!\!10}
%\bigg
\}
\end{equation}
\fi
%
% where $u(\theta_{ij}^{NHO},\theta_{ij}^{HOC'}) =
where $u_{ij} =
 \cos^2 \theta_{ij}^{NHO} \cos^2 \theta_{ij}^{HOC'}$ for $\pi/2  < \theta_{ij}^{NHO}$,
 $\theta_{ij}^{HOC'} < 3\pi/2$ and zero otherwise,
\cut{
u(\theta_{ij}^{NHO},\theta_{ij}^{HOC'}) & = & \left\{ \begin{array}{ll}
 \cos^2 \theta_{ij}^{NHO} \cos^2 \theta_{ij}^{HOC'} & \pi/2  < \theta_{ij}^{NHO},
 \theta_{ij}^{HOC'} < 3\pi/2 \\
0 & {\rm otherwise}
\end{array} \right.
\end{eqnarray}
}
%where
$r_{ij}^{HO}$ is the distance between $H(i)$ and $O(j)$,
$\theta_{ij}^{NHO}$ is the $N(i)H(i)O(j)$ angle and $\theta_{ij}^{HOC'}$
the $H(i)O(j)C'(j)$ angle.

%\paragraph{Hydrophobic, surface and related interactions.}
The hydrophobic interaction $E_{\rm hp}$ consists of two types of terms.
The first one is a pure radial 12-6
Lennard Jones potential that should take into account
all non-local, hydrophobic and other forces between the
$i$th amino acid $a_i$ and the $j$th $a_j$.
% $a_i=\{{\rm Ala},\ldots,{\rm Val}\}$.
Both $C_\alpha C_\alpha$ and
$RR$ interactions are included. The $C_\alpha C_\alpha$ interaction,
i.e.\ $\sum_{i>j} \epsilon^\alpha(a_i,a_j)
\left\{
5 \left(\frac{\sigma^\alpha(a_i,a_j)}{r_{ij}^{\alpha}} \right)^{\!\!12}
\!\!\!\!
-6 \left( \frac{\sigma^\alpha(a_i,a_j)}{r_{ij}^{\alpha}} \right)^{\!\!6}
\right\}$, where $r_{ij}^\alpha$  is the $C_\alpha(i)C_\alpha(j)$ distance,
is mainly introduced to model steric constraints. The second type of term, which
plays a minor role, is a surface energy term inspired by Refs.\
%\cite{TaLu-ScWo99,SiRuKoFoByBa99}
\cite{BaCASP4}
$
\sum^{R}_i \epsilon^{\rm surf}_i
f\left[ \frac{1}{s_i}
\sum^{R}_{j\neq i-1,i,i+1}
f \left(
\frac{  \sigma^{{\rm surf,}0}_j -r^R_{ij}}{\sigma^{\rm surf}_j} \right)
- \frac{b_i}{s_i}
\right]
%}
$,
\iffalse
\beqa
E_{\rm hp} &=& \sum_{i>j} \epsilon^\alpha(a_i,a_j)
\bigg\{
5 \bigg(\frac{\sigma^\alpha(a_i,a_j)}{r_{ij}^{\alpha}} \bigg)^{\!\!12}
\!\!\!\!
-6 \bigg( \frac{\sigma^\alpha(a_i,a_j)}{r_{ij}^{\alpha}} \bigg)^{\!\!6}
\bigg\}
\nonumber \\
\lefteqn{
+ \sum_{i>j} \epsilon^R(a_i,a_j)
\bigg\{
5 \bigg( \frac{\sigma^R(a_i,a_j)}{r_{ij}^R} \bigg)^{\!\!12}
\!\!\!\!
-6 \bigg( \frac{\sigma^R(a_i,a_j)}{r_{ij}^R} \bigg)^{\!\!6}
\bigg\}
} \nonumber \\
\lefteqn{+ \sum^{R}_i \epsilon^{\rm surf}_i
f\bigg[ \frac{1}{s_i}
\sum^{R}_{j\neq i-1,i,i+1}
f \bigg(
\frac{  \sigma^{{\rm surf,}0}_j -r^R_{ij}}{\sigma^{\rm surf}_j} \bigg)
- \frac{b_i}{s_i}
\bigg],
}
\eeqa
\fi
where
$f(x) \equiv 0.5(1+\tanh(x))\in [0,1]$ is a sigmoid non-linearity.
The surface energy term is chosen such that
$\epsilon^{\rm surf}_i=\epsilon^{\rm surf}(a_i)$
is the energy change induced by taking side chain $a_i$ from
being completely exposed to solvent to being completely buried.
The inner sum counts the number of neighboring amino-acids and
the adjustable parameters
$s_i,b_i,\sigma^{{\rm surf,}0}_i,\sigma^{\rm surf}_i$ set the relevant scale and bias for
burial of each amino acid.
The total number of adjustable parameters for the hydrophobic energy is
 2*2*(20*19/2+20)=840 for Lennard Jones and 5*20=100 for surface.

The temperature scale is arbitrary since the
Boltzmann weight only depends upon the product $\beta E$ and the scale of $E$ is set during
training. In the test below we choose the folding 'temperature' $1/\beta_{\rm fold} = 0.1$ for
all training examples and set the
initial parameter values to be on the same scale.
The initial values are chosen such that
all amino-acids have the same parameter values,
except for different
surface energy terms ($\epsilon^{\rm surf}(aa_i)$).

To get the learning algorithm eq.\
(\ref{Boltzmann_update}) to work properly we need reliable estimates of
 thermodynamic averages. To achieve this we use parallel tempering \cite{GeTh95},
where the system is simulated independently at a number
$N_{\rm temp}$ of different
temperatures.  In this study $N_{\rm temp}=15$ ranging from
$T_{\rm min}=0.1$ to $T_{\rm max}=0.8$ with equal spacing on a log-scale.
Once every cycle, where
a cycle consists of $N_{\rm conf}$ elementary conformation updates, the temperature of
two random systems (adjacent in temperature)
%\footnote{To get a higher acceptance rate, normally only pairs of
%systems adjacent in temperature are selected.}
are exchanged with the Metropolis probability
$P({\rm accept}) = \min(1,\exp(\Delta \beta \Delta E))$ where $\Delta E$ and  $\Delta \beta$
are the energy and inverse temperature difference between the two systems respectively.
In our case we use $N_{\rm conf}=40$ of which one quarter are pivot moves (rotation around a
random torsional angle) and the rest are 'local' moves which is defined as choosing two torsional
angles next to the same peptide bond (i.e.\ $\psi_i$ and $\phi_{i+1}$) and rotate them
opposite angles.
%For small rotation angles,
%the local move will (to a good approximation) only rotate the
%peptide group
%and leave the conformation of the rest of the structure almost unaltered.
We
collect statistics for $10^4$ cycles between each update of the adjustable parameters. We choose the folding temperature to be the minimum temperature $T_{\rm fold} =T_{\rm min}$ and thus only use the statistics for this temperature to
update the parameters.
To ensure sampling of the native conformations, a number of the $N_{\rm temp}$ systems are initialized in the native state. The remaining systems are
initialized in different
low energy states found in the previous update to ensure fast focus on relevant
regions of conformation space. Subsequent long runs starting from coil
confirm that this procedure is sufficiently close to generate
equilibrated samples.
As an alternative to the batch update rule eq.\ (\ref{Boltzmann_update}), one can use an online
version where the parameters are updated using a single training example at a time.
In this study, we use an intermediate approach, where we update the parameters using three
batches each with one third of the example. The results presented below are obtained using
approximately 500 parameter updates. The training set consists of a small set of
 24 protein fragments (or peptides)
of length 11-14 of mainly $\alpha$-helices and 
$\beta$-turns\footnote{Protein fragments used, PDB-code and amino acids:
      1ALC 21-32,  1BGS 10-22, 1BGS 88-98,  1FKF 27-38, 1HGF 100-113,
      1HRC 91-102, 1I1B 101-112,  1MBC 6-17,  1MBC 29-40,  1MBC 99-111,
      1MBC 131-142, 1PGA 43-54,  1UBQ 3-15, 2I1B 69-82, 2I1B 103-112,
      2MHR 51-62,   2MHR 67-78,  2MHR 102-113, 2PCY 18-29, 3LZM 24-35,
      3LZM 99-111, 4PTI 22-33,  5CYT 88-101, 7RSA 2-13.
All sequences and structures are listed in Ref.\ \cite{PeMo97}.}. They have 
been suggested to adopt their native structure
even as fragments \cite{MoUn91}.
Running the training on 8 processors on a Silicon Graphics Origin 3000
computer, 500 parameter updates take approximately 2 CPU weeks.

\begin{figure}
\centerline{\includegraphics[width=0.4\textwidth]{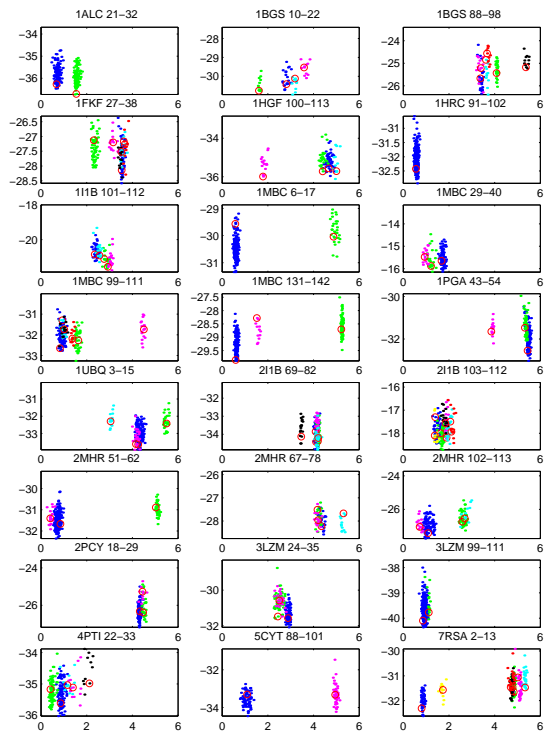}}
%\vspace{22cm}
%\special{psfile=decoy_tmp_ordered_1_24.ps hscale=100 vscale=100 hoffset=-87 voffset=-90 angle=0}
\caption{Color online. Decoy plots -- energy versus RMSD (\AA). Cluster RMSD cut-off is 0.5 \AA.
Cluster centers are marked with a circle. Blue codes for the largest cluster.}
\label{fig:decoy}
\end{figure}

We tested the final potential by initializing the 24 training
sequences in random coil. After an initial equilibration,
conformations were saved with fixed intervals at the folding temperature
$T_{\rm fold}=0.1$ in a long test run. These sampled conformations are
called decoys below.
Some results from the test
run are shown in Figure \ref{fig:decoy}.
The decoys are clustered by introducing a
RMSD cut-off of 0.5 {\AA} and assigning as the first cluster center the decoy
with the most neighbors within the cut-off.
The clustering is not very sensitive
to the specific choice of the cut-off.
We remove these decoys and repeat
the procedure until all decoys have been assigned to a cluster. The number of
decoys in the cluster is directly related to the free energy
$F({\rm clus} \ i) = - T \ln P({\rm clus} \ i)$, where the probability
$P({\rm clus} \ i)$ of
cluster $i$ is the number of decoys in cluster $i$ divided
by the total number of decoys.
The decoy plots reveal a complex
free energy landscape with competing minima. In a few cases
free energy minima both with small and large RMSD to the native fold exist
simultaneously. In the subsequent analysis the cluster center of the
cluster with the lowest free energy was chosen as the predicted
fold.

\begin{figure}
\centerline{\includegraphics[width=0.275\textwidth]{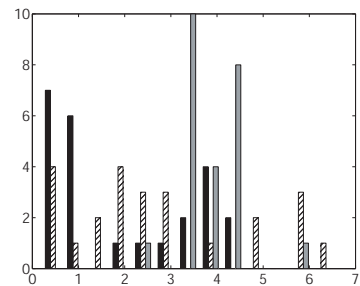}}
%\vspace{7cm}
%\special{psfile=histogram.ps hscale=45 vscale=45 hoffset=67.5 voffset=-90 angle=0}
\caption{The RMSD histogram for the minimum free energy conformations
for the 24 peptides after training (black), the results for
an all-atom potential \protect\cite{PeMo97} (hatched) 
and prior to training (gray).}
\label{fig:histogram}
\end{figure}

The performance on the training set is summarized in Figure
\ref{fig:histogram}. The trained potential is compared to the initial
essentially homo-polymer potential and the results of folding with
an all-atom potential from \cite{PeMo97}. A clear improvement over the initial
potential is seen, indicating that the training process actually works.
The results are also comparable to the all-atom potential,
%although the
%comparison is not entirely unbiased in this case since the all-atom
%potential is not tuned specifically for the training peptides. The comparison
%is interesting anyway because it 
which shows that our approach is
comparable to an all-atom potential that requires many human expert
man hours to derive. Probing the significance of the folding
temperature by performing the test run at a lower temperature
$T=0.025<T_{\rm fold}$ shows that
% The decoy plot is shown in the Supporting Material.
%As expected, the clusters are more narrow in RMSD and
the overall
performance---in terms of RMSD for the largest cluster---is
significantly worse.
%Only in one case (1HGF 100-113)
%a lower RMSD minimum free energy cluster is found.

\begin{figure}
\centerline{\includegraphics[width=0.40\textwidth]{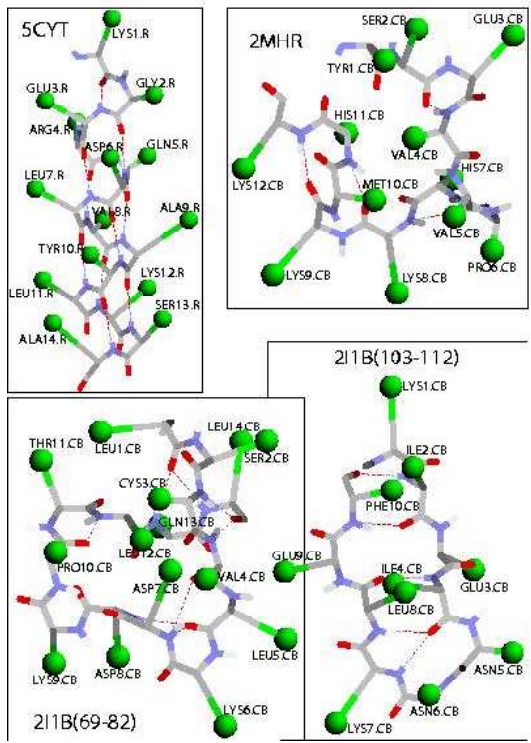}}
\caption{Color online. Representative decoys from the lowest free energy cluster for
four peptides from top left to right (ID, RMSD from native, native
structure type):
5CYT 88-101, 1.04{\AA}, $\alpha$-helix;
2MHR 67-78, 4.74{\AA}, $\alpha$-helix;
2I1B 69-82, 4.4{\AA}, $\beta$-turn and
2I1B 103-112, 1.63{\AA}, $\beta$-turn .}
\label{fig:structures}
\end{figure}

To get an understanding of the successes and failures of the potential we
have visualized low energy structures (see fig.\ \ref{fig:structures}) and
made Ramachandran plots for the amino acids.
The successful predictions are very native-like
making the same hydrogen bonds as the native structure.
However, some of the side chains are very close together.
Although a side chain should be regarded as 'effective'
with degrees of freedom averaged out and not as an atom,
the small distance means that the
characteristic separation in the Lennard Jones potential is small and will
be sensitive to small changes in the distance between the side chains.
The structure for some of the
failures are not 'protein-like' and some of the amino acids are not reproducing
the Ramachandran behavior found for real proteins.

These findings
show that the principle works, however,
it is clear that the potential function model can be improved in many ways.
One of the great advantages of the method is that many terms can
be added and if they do not work well their weight would end up
being very low. However, it should be kept in mind that the better
the starting point, the more likely it is to reach a reasonable
parameter set.
%Inspecting the solution found by the algorithm
The test suggests
that the representation of side chains in the potential with
just one pseudo-atom and a fixed angle is too crude. One
remedy is to make the side chain model more realistic, e.g.\ by introducing an
explicit  $C_\beta$ atom. This would probably make the Ramachandran behavior
'protein-like' and remove some of the false minima the model is currently
struggling with. It is also possible to go in the opposite direction and use
%a simple hopefully less rugged potential function, but with 
a more restricted conformational
search space, e.g.\ by only sampling experimentally observed Ramachandran angles
or %by 
using an I-Sites library to generate conformations
\cite{BaCASP4}. %,ChRoBoBa03}.
The two different views are complementary and the results of the CASP
exercise has shown that it is important
to pursue both to generate good ab initio predictions \cite{BaCASP4}.

The ultimate goal of optimizing potentials is to obtain
reasonable predictions for sequences not in the
training set (generalization).
Preliminary runs on such test sequences show poor generalization,
which is primarily a result of the small training set.
%To reliable
%estimate all pair potentials we need each pair of amino acids
%to appear at least order of 10 times in training set. 
%For the current training set most but not all amino acids 
%pairs were present.
It is therefore important to now scale up to a more
realistic size using more and longer sequences. 
We are currently working on ways to speed up the whole
process to achieve this goal.
%We are currently investigating distributing
%the computation and speeding up the individual parts of the
%process as well as optimizing the Monte Carlo method.
%
%The difficulty in doing so is that
%the computation time for equilibration of a single sequence
%is believed to be between quadratic and cubic in the sequence length
%depending on the implementation. We are working on distributing
%the computation and speeding up the individual parts of the
%process as well as optimizing the Monte Carlo method.
%We also believe that the method is likely to work even
%if the proteins are not fully equilibrated, which could potentially
%speed it up significantly.

More details about parameter settings, data sets and
results can be found at {\tt www.imm.dtu.dk/$\sim$owi/}.

%\section*{Acknowledgements}

{\it Acknowledgments.} This work was sponsored by a grant to the Center for Biological
Sequence Analysis (S{\o}ren Brunak) from the Danish National Research
Foundation.

\iffalse
\begin{figure}[ht]
\vspace{7cm}
\includegraphics{b_041201_1GB1_update_1_min_E_native_fit_to_pdb.ps}
\caption{Protein G, 56 residues, low energy structure fit to experimental structure, RMSD=1.9 \AA, visualized with MolMol \protect\cite{KoBiWu96}.}
\label{fig:proteinG}
\end{figure}
\fi

\newcommand{\etal}{\textit{et al.}}

\end{document}